\documentclass[12pt,letterpaper]{article}
\usepackage[pdftex]{graphicx,color}
\input{epsf}
\usepackage{amsmath,amssymb}
\usepackage[dvips,letterpaper,text={6.5in,9in}]{geometry}
\usepackage{fancyhdr}
\usepackage{verbatim}

\newcommand\ltap{\
  \raise.3ex\hbox{$<$\kern-.75em\lower1ex\hbox{$\sim$}}\ }
\newcommand\gtap{\
  \raise.3ex\hbox{$>$\kern-.75em\lower1ex\hbox{$\sim$}}\ }

\newcommand\simge{\mathrel{%
   \rlap{\raise 0.511ex \hbox{$>$}}{\lower 0.511ex \hbox{$\sim$}}}}
\newcommand\simle{\mathrel{
   \rlap{\raise 0.511ex \hbox{$<$}}{\lower 0.511ex \hbox{$\sim$}}}}

\newcommand{\slashchar}[1]%
        {\kern .25em\raise.18ex\hbox{$/$}\kern-.75em #1}
\def\lsim{\mathrel{\raise.3ex\hbox{$<$\kern-.75em\lower1ex\hbox{$\sim$}}}}
\def\gsim{\mathrel{\raise.3ex\hbox{$>$\kern-.75em\lower1ex\hbox{$\sim$}}}}
\newcommand{\bs}{\boldsymbol}
\newcommand{\Tr}{{\rm Tr}}

\newcommand\CG{{\cal G}}

\newcommand\CK{{\cal K}}
\newcommand\CL{{\cal L}}

\newcommand\CO{{\cal O}}

\newcommand\be{\begin{equation}}
\newcommand\ee{\end{equation}}
\newcommand\bea{\begin{eqnarray}}
\newcommand\eea{\end{eqnarray}}
\newcommand\ba{\begin{array}}
\newcommand\ea{\end{array}}
\newcommand\nn{\nonumber}

\newcommand{\half}{\ensuremath{\frac{1}{2}}}
\newcommand{\thalf}{\textstyle{\frac{1}{2}}}

\newcommand{\tfourth}{\textstyle{\frac{1}{4}}}

\newcommand{\thw}{\ensuremath{\theta_W}}

\newcommand\dagg{\dagger}

\newcommand\ra{\rightarrow}

\newcommand\mev{{\rm MeV}}
\newcommand\gev{{\rm GeV}}
\newcommand\tev{{\rm TeV}}

\newcommand\ifb{{\rm fb}^{-1}}

\newcommand\cm{{\rm cm}}
\newcommand\ellm{\ell^-}

\newcommand\ellp{\ell^+}

\newcommand\Ntc{N_{TC}}

\newcommand\atc{\alpha_{TC}}

\newcommand\Leff{{\cal L}_{\rm eff}}
\newcommand\Lsig{{\cal L}_{\Sigma}}
\newcommand\LFF{{\cal L}_{\rm gauge}}
\newcommand\LWZW{{\cal L}_{\rm WZW}}
\newcommand\Lff{{\cal L}_{\bar f f}}
\newcommand\Lpifbf{{\cal L}_{\tpi \bar f f}}
\newcommand\grpp{g_{\rho_T\pi_T\pi_T}}

\newcommand\tom{\omega_{T}}
\newcommand\tro{\rho_{T}}

\newcommand\ta{a_T}

\newcommand\st{\sigma_T}

\newcommand\tpi{\pi_T}
\newcommand\tpipm{\pi_T^\pm}
\newcommand\tpimp{\pi_T^\mp}
\newcommand\tpip{\pi_T^+}
\newcommand\tpim{\pi_T^-}
\newcommand\tpiz{\pi_T^0}
\newcommand\tpipr{\pi_T^{0 \prime}}

\newcommand\jets{{\rm jets}}

\begin{document}
\title{
\vskip -15mm
\begin{flushright}
\vskip -15mm
{\small FERMILAB-Pub-10-442-PPD-T\\
  LAPTH-044/10\\
  CERN-PH-TH-2010-215\\
}
\vskip 5mm
\end{flushright}
{\Large{\bf A Light Scalar in Low-Scale Technicolor}}\\
} \author{
  {\large Antonio Delgado$^{1}$\thanks{antonio.delgado@nd.edu} ,\,
  Kenneth Lane$^{2,3}$\thanks{lane@physics.bu.edu} \, and
  Adam Martin$^{4}$\thanks{aomartin@fnal.gov}}\\
{\large $^{1}$Department of Physics, University of Notre Dame}\\
{\large Notre Dame, Indiana 46556}\\
{\large $^{2}$Department of Physics, Boston University}\\
{\large 590 Commonwealth Avenue, Boston, Massachusetts 02215}\\
{\large $^{3}$LAPTH\thanks{Laboratoire de Physique Th\'eorique
    d'Annecy-le-Vieux, UMR5108}\,, Universit\'e de Savoie, CNRS} \\
{\large B.P.~110, F-74941, Annecy-le-Vieux Cedex, France}\\
{\large {$^{4}$}Fermi National Accelerator Laboratory}\\
{\large Batavia, Illinois 60510}\\
}
\maketitle

\begin{abstract}
  In addition to the narrow spin-one resonances $\tro$, $\tom$ and $\ta$
  occurring in low-scale technicolor, there will be relatively narrow scalars
  in the mass range 200 to 600--700~GeV. We study the lightest isoscalar
  state, $\st$. In several important respects it is like a heavy Higgs boson
  with a small vev. It may be discoverable with high luminosity at the LHC
  where it is produced via weak boson fusion and likely has substantial $W^+
  W^-$ and $Z^0Z^0$ decay modes.

\end{abstract}


\newpage

{\em {1. \underbar {Introduction}}} \,\, Walking
technicolor~\cite{Holdom:1981rm, Appelquist:1986an,Yamawaki:1986zg,
  Akiba:1986rr} is an asymptotically free gauge theory whose coupling $\atc$
runs very slowly for 100s, perhaps 1000s, of TeV above the electroweak
breaking scale of a few 100~GeV. This is necessary so that extended
technicolor (ETC) can generate sizable quark and lepton masses while
suppressing otherwise fatal flavor-changing neutral current
interactions~\cite{Eichten:1979ah}. A suitable walking $\atc$ occurs if the
critical coupling for chiral symmetry breaking lies just {\em below} an
(approximate) infrared fixed point~\cite{Lane:1991qh, Appelquist:1997fp}.
This requires a large number $N_D \gg 1$ of technifermion doublets in the
fundamental representation of the TC gauge group {\em or} a few doublets in
higher-dimensional representations~\cite{Lane:1989ej,Dietrich:2005wk}. In the
latter case, the constraints on ETC representations~\cite{Eichten:1979ah}
almost always imply other technifermions in the fundamental representation as
well. Thus, it is expected that there are technifermions whose technipions
($\tpi$) have a decay constant $F_1^2 \ll F_\pi^2 = (246\,\gev)^2$. This
implies that bound states of the lightest technifermion doublet $(T_U,T_D)$
have masses well below a TeV --- greater than the experimental lower limit of
225--250~GeV~\cite{Abazov:2006iq, Aaltonen:2009jb} and probably not more than
600--700~GeV. We refer to this as {\em low-scale technicolor} (LSTC). The
most experimentally accessible bound states are the technivectors $V_T =
\tro(I^G J^{PC} = 1^+1^{--})$, $\tom(0^-1^{--})$ and $\ta(1^-1^{++})$, all of
which may be produced as $s$-channel resonances of the Drell-Yan process in
hadron and lepton colliders. The technipions $\tpi(1^-0^{-+})$ may be
accessed through $V_T$ decays.\footnote{The isoscalar axial-vector
  $f_T(0^+1^{++})$ can be produced via weak vector boson fusion but not by
  the much stronger Drell-Yan process. We will not consider it further in
  this paper.}  A central assumption of LSTC is that the lightest
technihadrons may be treated in isolation, without significant mixing or
other interference from higher-mass states. In a model with $N_D$ equivalent
technifermion doublets, this requires that ETC leaves just one of them
significantly lighter than the others. We also assume that the lightest
technifermions are ordinary $SU(3)$-color singlets. For a more extensive
discussion of LSTC, see Ref.~\cite{Lane:2009ct}.

Walking technicolor has another important phenomenological consequence: It
enhances $M_{\tpi}$ relative to the lightest $M_{\tro}$ so that the
all-$\tpi$ decay channels of $\tro$, $\tom$ and $\ta$ are
closed~\cite{Lane:1989ej}. The light $V_T$ then are {\em very} narrow,
$\simle 1\,\gev$, with principal decays to $\tpi+W$ or $Z$, a pair of
electroweak bosons (including one photon), and fermion-antifermion
pairs~\cite{Lane:2002sm,Eichten:2007sx}. Many of these provide striking
signatures, visible above backgrounds, within a limited mass range at the
Tevatron and probably up to 600--700~GeV at the
LHC~\cite{Brooijmans:2008se,Brooijmans:2010tn}.

A walking $\atc$ also invalidates~\cite{Lane:1993wz,Lane:1994pg} the
QCD-based assumptions made to estimate the $S$-parameter for technicolor
models~\cite{Peskin:1990zt,Golden:1990ig, Holdom:1990tc,Altarelli:1991fk}.
Further, it has been suggested that walking may cause $\tro$ and $\ta$ to be
closer in mass than their QCD counterparts and to have approximately equal
couplings to the vector and axial-vector parts of the electroweak
currents~\cite{Appelquist:1998xf, Appelquist:1999dq, Hirn:2006nt,
  Hirn:2006wg,Foadi:2007ue,Elander:2009pk, Elander:2010wd}. That would eliminate the low-scale contribution
to $S$. Determining the viability of this conjecture is the object of several
recent papers employing lattice-gauge techniques~\cite{Catterall:2007yx,
  Appelquist:2007hu,Appelquist:2010xv}. Consequently, LSTC phenomenology now
assumes that $M_{\tro} \cong M_{\tom} \simle M_{\ta} < 2 M_{\tpi}$, with
isospin-symmetric masses.

The main point of this paper is that, given this pattern of LSTC masses,
there will be scalars $\sigma_0(0^+0^{++})$ and ${\bs \sigma_1}(1^-0^{++})$,
analogs of the $f_0$ and $a_0$ of QCD. As with the technivectors, we expect
that ETC enhancements to their masses are less important than they are for
technipions, so that $M_{\tro} \cong M_{\tom} \simle M_{\sigma_{0,1}} \simle
M_{\ta} \simle 2M_{\tpi}$. Thus, these scalars are also relatively narrow
(compared to a standard-model Higgs boson of the same mass). Like
technipions, their couplings to $\bar ff$ are induced by ETC boson exchange.
As for the $\tpi$, they are of order $m_f/F_1$ --- except for the top quark
because ETC produces no more than 5--10~GeV of the top's
mass~\cite{Hill:1994hp, Lane:2009ct}.  Therefore, so long as the $\sigma_0$'s
constituent fermions are color singlets, its main production mechanism at the
LHC is weak vector boson fusion, not gluon fusion.\footnote{The $\sigma_0$
  mixing with the light $\bar t t$ scalar that would be expected in a
  topcolor-assisted technicolor model is small because it is proportional to
  the $\sigma_0\bar tt$ coupling induced by ETC. Also, if QCD is any guide,
  $0^+0^{++}$ techni-glueballs are considerable heavier than $\sigma_0$ so
  that mixing between these scalars is not significant.}  The ${\bs
  \sigma_1}$ is not produced by either mechanism. Therefore, in this Letter
we concentrate on $\sigma_0$, which we refer to as $\st$ from now on.  Its
principal production and decay modes are similar to those of a heavy Higgs
boson in a type-I two-Higgs-doublet model.  It has a small vev, couples to
the $W$ and $Z$, but only weakly to quarks and leptons. The LHC with
$\sqrt{s} = 14\,\tev$ can produce $\sim 1$~to~100 $\st$ per $\ifb$ in the
mass range 200--700~GeV and decaying to $WW$ and $ZZ$ in all-leptonic or
semileptonic channels. However, we shall see that several $100\,\ifb$ will
likely be required to discover $\st$ in these modes.

A light scalar with vacuum quantum numbers in walking technicolor has been
proposed by a number of authors; see, e.g., Refs.~\cite{Yamawaki:1986zg,
  Holdom:1987yu,Dietrich:2005jn,Galloway:2010bp,Appelquist:2010gy,Yamawaki:2010ms}.
Usually, this is a ``techni-dilaton'', a pseudo-Goldstone boson argued to
arise as a consequence of spontaneous breaking of the theory's approximate
conformal invariance. We do not believe that $\st$ is a techni-dilaton. The
main phenomenological difference between the two is in their couplings to
matter and gauge fields. The techni-dilaton couples as $F_\pi/F_c$, where
$F_c \gg F_\pi$ is the scale at which the conformal symmetry is broken while,
as we see next, $\st$ couples as $F_1/F_\pi$. Although these couplings may be
numerically similar, they have different origins. Furthermore, the
technihadron partners of $\st$ and its place in their spectrum is specific to
the version of LSTC considered here.

{\em 2. {\underbar{Effective Lagrangian for $\st$ in Low-Scale
      Technicolor}}}\,\, In Ref.~\cite{Lane:2009ct} two of us constructed an
effective Lagrangian for LSTC. It describes the interactions at energies
$\simle M_{\tro}$ of the lowest-lying technihadrons. A principal motivation
for constructing $\Leff$ was to provide a consistent treatment of the weak
bosons, including the longitudinal $W_L$ and $Z_L$, which are common products
of $V_T$ decays. The $V_T$ are included using the hidden local symmetry (HLS)
formalism of Bando, {\em et al.}~\cite{Bando:1984ej, Bando:1987br}. The
Lagrangian is based on $\CG = SU(2)\otimes U(1)\otimes U(2)_L \otimes
U(2)_R$, where $SU(2)\otimes U(1)$ is the electroweak gauge group and $U(2)_L
\otimes U(2)_R$ is the HLS gauge group.

To describe the lightest technihadrons and mock up the heavier TC
states contributing most to electroweak symmetry breaking (i.e., the
isovector technipions of the other $N_D - 1$ technifermion doublets or the
higher-scale states of a two-scale TC model), and to break all the gauge
symmetries down to electromagnetic $U(1)$, we used nonlinear  $\Sigma$-model
fields $\Sigma_2$, $\xi_L$, $\xi_R$, $\xi_M$ and $\Sigma_1 \equiv \xi_L \xi_M
\xi_R$, with covariant derivatives
\bea\label{eq:covderivs}
D_\mu \xi_L &=& \partial_\mu\xi_L -i(g\bs{t}\cdot\bs{W}_\mu +g' y_1 t_0
B_\mu) \xi_L + ig_T \xi_L\, t\cdot L_\mu\nn\\
D_\mu \xi_M &=& \partial_\mu\xi_M  -ig_T (t \cdot L_\mu\, \xi_M - \xi_M\,  t
\cdot R_\mu) \nn\\
D_\mu \xi_R &=& \partial_\mu\xi_R -ig_T t \cdot R_\mu\, \xi_R + ig' \xi_R (t_3
+ y_1 t_0)B_\mu \nn\\
D_\mu \Sigma_{1,2} &=& \partial_\mu\Sigma_{1,2} -ig \bs{t}\cdot\bs{W}_\mu
\Sigma_{1,2} + ig' \Sigma_{1,2} t_3 B_\mu\,,
\eea
where $t\cdot L_\mu = \sum_{\alpha=0}^3 t_\alpha L_\mu^\alpha$ and $\bs{t} =
{\half}\bs{\tau}$, $t_0 = {\half}\bs{1}$. The HLS gauge coupling $g_L = g_R =
g_T$ reflects the parity invariance of TC interactions and the expectation
that $I=0,1$ technivectors are nearly degenerate. The hypercharge $y_1 = Q_U
+ Q_D$ is the sum of electric charges of $T_U$ and $T_D$.  The field
$\Sigma_2$ contains the technipions that get absorbed by the $W$ and $Z$
bosons. We represent them as an isotriplet of $F_2$-scale Goldstone bosons,
where $F_2^2 = (F_\pi \cos\chi)^2 \gg F_1^2$, and $\chi$ is a mixing angle
with, e.g., $\sin\chi \simeq 1/\sqrt{N_D}$ in an $N_D$ doublet model.

Although $\st$ was not included in the nonlinear fields in
Ref.~\cite{Lane:2009ct}, it is easy to incorporate. In the unitary gauge, in
which $\Sigma_2, \xi_L, \xi_R \ra 1$ and $\Sigma_1 = \xi_M$, we write $F_1
\Sigma_1 = (\st + F_1) E$ where $E = \exp{(2i{\bs t}\cdot{\bs
    \pi_T})/F_1}$.\footnote{These $\st$ and ${\bs \pi_T}$ fields are not yet
  canonically normalized. Also, we have assumed the isoscalar $\tpipr$ to be
  much heavier than $\tro$ and integrated it out. See Ref.~\cite{Lane:2009ct}
  for discussions of these points.} We do not consider other light scalars to
arise from $\xi_{L,R}$ because they are not expected in the low-lying
spectrum of $\bar T T$-hadrons. The complete effective Lagrangian is
\be\label{eq:Leff}
\Leff = \Lsig + \LFF + \Lff + \LWZW + \CL_{M^2} + \Lpifbf\,,
\ee
where, in unitary gauge,
\bea\label{eq:Lsigtwo}
\Lsig = && {\tfourth} F_2^2\, \Tr\bigl|g\bs{t}\cdot\bs{W}_\mu - g' t_3
            B_\mu\bigr|^2 + \thalf(a+c)(\partial_\mu\st)^2 \nn\\
      &&\,+\, {\tfourth} (\st+F_1)^2 \Bigl\{a\Tr\bigl|\partial_\mu E
              -i(g\bs{t}\cdot\bs{W}_\mu E - g'E t_3 B_\mu)\bigr|^2 \nn\\
      && \,+\, b\Bigl[\Tr\bigl|g\bs{t}\cdot\bs{W}_\mu + g'y_1 t_0 B_\mu
                   - g_T t\cdot L_\mu \bigr|^2 +\Tr\bigl|g'(t_3 + y_1 t_0)B_\mu
                - g_T t\cdot R_\mu\bigr|^2 \Bigr] \nn\\
        && \,+\, c\, \Tr\bigl|\partial_\mu E + ig_T( E t\cdot R_\mu
        - t\cdot L_\mu  E)\bigr|^2 \\
        && \,+\, d\, \Tr\bigl[(g E^\dagg \bs{t}\cdot\bs{W}_\mu -
        g' t_3  E^\dagg B_\mu + g_T(t\cdot R_\mu E^\dagg -
         E^\dagg t\cdot L_\mu)) \nn\\
        &&\qquad\,\, \times (\partial_\mu E + ig_T( E t\cdot
        R_\mu - t\cdot L_\mu E))\bigr]\Bigr\} \nn\\
        && \,-\frac{if}{2g_T F_1^2}\, (\st + F_1)^2 \nn\\
        && \qquad\,\, \times \Bigr\{\Tr\bigl[(\partial_\mu E +
            ig_T( E t\cdot R_\mu - t\cdot L_\mu E)) E^\dagg
          (\partial_\nu E + ig_T( E t\cdot R_\nu - t\cdot L_\nu E)) E^\dagg
            t\cdot L^{\mu\nu} \nn\\
        &&\qquad\,\, +  E^\dagg(\partial_\mu E +
            ig_T( E t\cdot R_\mu - t\cdot L_\mu  E))
                        E^\dagg(\partial_\nu E +
            ig_T( E t\cdot R_\nu - t\cdot L_\nu  E))t\cdot R^{\mu\nu}
            \bigr] \Bigr\} \,.\nn
\eea
A ``simplicity principle'' was adopted in writing $\Lsig$: only the
lowest-dimension operators needed to describe the experimentally important
LSTC processes --- mainly $V_T$ two-body production and decay vertices ---
were kept. The dimensionless parameters $a,\dots,f$ are nominally of
$\CO(1)$. The gauge-field Lagrangian $\LFF$ has the standard form; $\Lff$ is
the usual coupling of quarks and leptons to the $SU(2)\otimes U(1)$ gauge
bosons {\em only}; $\CL_{M^2}$ describes $\tpi$ and $\st$ masses and
$\Lpifbf$ the $\tpi \bar ff$ couplings. The Wess-Zumino-Witten interaction
$\LWZW$ reflects the anomalous global {\em and} HLS gauge symmetries of the
underlying theory~\cite{Harvey:2007ca}. It is essential for describing
radiative decays of $\tro$ and $\tom$ as well as $\tpiz \ra \gamma\gamma$.
Processes computed in tree approximation, including production and scattering
of $W_L$ and $Z_L$, behave at high energies, $\sqrt{s} \gg M_{\tro}$, as they
do in the standard model without a Higgs boson.\footnote{By itself, this
  $\st$ does not alter this situation much because of its small $\CO(F_1)$
  coupling to $WW$.} The masses of the $W$, $Z$ and $V_T$, with
$M_W/M_Z\cos\thw = 1$ and $M_{\tom} = M_{\tro} + \CO((gy_1\sin\thw/g_T)^2)$,
follow from $\Lsig$. See Ref~\cite{Lane:2009ct} for details.


The mixing angle $\chi$ characterizing the contribution of the low
$F_1$-scale to electroweak symmetry breaking is
\be\label{eq:sinchi}
\sin\chi \simeq F_1/F_\pi\,,
\ee
where $F_\pi = \sqrt{F_2^2 + AF_1^2/B} = 246\,\gev$.

In the $U(2)_V$ limit in which $M_{\tom} = M_{\tro}$, all quantities of
phenomenological interest can be expressed in terms of $M_{\tro}$, $M_{\ta}$,
$F_\pi$, the number of technicolors $\Ntc$, $\sin\chi$, $y_1 = Q_U+Q_D$, and
three mass parameters --- $M_{V_1}$, $M_{A_1}$, $M_{A_2}$ --- that control
the strength of dimension-five operators involved in decays of technivectors
to photons and transversely-polarized weak bosons.\footnote{We assume for
  simplicity that the TC gauge group is $SU(\Ntc)$ and the technifermions
  transform as the fundamental ${\bs \Ntc}$. This affects WZW interactions
  whose dimension-five operators are $\propto 1/M_{V_1}$.} In particular, the
effective coupling $\grpp$ for $\tro \ra \tpi\tpi$, $\tpi W_L$ and $W_L W_L$
is~\cite{Brooijmans:2010tn}
\be\label{eq:massinputs}
\grpp = \frac{M_{\tro}^2}{\sqrt{2}g_T (F_\pi\sin\chi)^2}\left[1 +
  (f-1)\frac{M_{A_2}^2}{M_{A_1}^2}\right]\,,
\ee
in which
\be\label{eq:gTf}
g_T = \frac{16\sqrt{2}\pi^2 M_{A_1}F_\pi\sin\chi}{\Ntc M_{V_1}(M_{A_1} +
  M_{A_2})}\,, \qquad
f = \frac{(4\pi M_{A_1}F_\pi\sin\chi)^2}{\Ntc M_{V_1} M_{A_2}^2(M_{A_1} +
  M_{A_2})}\,.\nn
\ee
This ability to express unknown couplings in terms of the natural inputs of
LSTC is convenient. In the commonly used case, $M_{V_i} = M_{A_i} =
M_{\tro}$\footnote{This is motivated by the mass controlling ordinary
  $\rho,\omega \ra \gamma \pi^0$, namely, $M_{V_1(QCD)} \simeq 700\,\mev
  \simeq M_\rho$.}, $\grpp$ is given by the KSFR-like relation $\grpp =
M_{\tro}/2F_\pi\sin\chi$, while $g_T = 8\sqrt{2}\pi^2 F_\pi\sin\chi/\Ntc
M_{\tro}$. For light $\tro$, this relation has the consequence that $\tro \ra
W_L\tpi,\,W_L W_L$ are significantly suppressed (and radiative $\tro$ decay
branching ratios enhanced!) relative to those calculated using {\sc
  Pythia}~\cite{Sjostrand:2006za} where the default, scaled from QCD, is $g_T
= \grpp = \sqrt{4\pi(2.16)(3/\Ntc)}$.

At this point, we mention that the contribution of the $F_1$-scale $\tro$ and
$\ta$ to the $S$-parameter is~\cite{Lane:2009ct}
\be\label{Sone}
S_1(\tro,\ta)= \frac{8\pi}{g_T^2} \left(1 - \frac{M_{A_2}^2}{M_{A_1}^2}
\right)\,.
\ee
This can be made as small as desired, and it vanishes in the commonly used
case mentioned above. The contribution of $\st$ and the $\tpi$ is
\be\label{Ssztpi}
S_1(\st,\tpi) = \frac{1}{2\pi} \int_0^1 dx \, x(1-x)
\ln\left[\frac{M_{\tpi}^2 x + M_{\st}^2 (1-x)}{M_{\tpi}^2}\right]\,.
\ee
This is positive for $M_{\st} > M_{\tpi}$, but also quite small,
$\CO(10^{-2})$, for reasonable choices of these masses. Because of the
built-in isospin symmetry, $T_1 =0$ and other precision parameters are
negligibly small~\cite{Lane:2009ct}.

{\em 3. {\underbar {$\st$ at the LHC}}} \,\, Since $\st$ couples weakly to
$\bar t t$, its major production modes at the LHC are weak vector boson
fusion (VBF), $W^+W^-$ and $Z^0 Z^0 \ra \st$, and associated production,
$W^\pm, Z^0 \ra \st + W^\pm, Z^0$.  Its principal decay modes are $\st \ra
W^+W^-, \, Z^0 Z^0,\, W^\pm \tpimp,\, Z^0\tpiz$. The Lagrangian describing these
couplings comes from $\Lsig$ and is given by
\bea\label{eq:szprod}
\CL_{\st} &=& \CK \,\st \biggl[\tfourth F_\pi\sin\chi\biggl(2 g^2 W^{+\mu}
    W^-_\mu + (g^2 + g^{\prime 2})Z^\mu Z_\mu \biggr)\nn\\
&&\qquad - \cos\chi\biggl(g(W^+_\mu \partial^\mu \tpim + W^-_\mu \partial^\mu
\tpip)
  + \sqrt{g^2 + g^{\prime 2}}\,Z_\mu\, \partial^\mu \tpiz  \biggr) \biggr]\,,
\eea
where
\bea\label{eq:CK}
\CK &=& \frac{(4\pi M_{A_1} F_\pi \sin\chi)^4 - (\Ntc M_{\tro} M_{V_1}
  (M_{A_1}^2  - M_{A_2}^2))^2}
{(4\pi M_{A_1} F_\pi \sin\chi)^2 \Bigl[(4\pi M_{A_1}F_\pi\sin\chi)^4 +
 (\Ntc M_{\ta} M_{V_1} (M_{A_1}^2 - M_{A_2}^2))^2\Bigr]^{\thalf}}\nn\\
&\ra& 1 \,\,{\rm as}\,\, M_{A_1} - M_{A_2} \ra 0\,.
\eea
In the limit $\CK = 1$, $\CL_{\st}$ has the same form as the corresponding
interaction of a Higgs boson with vev $v_1 = F_\pi \sin\chi$. This lowers the
$\st$ production and decay rates by $\sin^2\chi$, nominally an order of
magnitude for the value $\sin\chi \simeq 1/3$ assumed in most LSTC
studies. Including the $W/Z \tpi$ decay modes, which a Higgs boson does not
have, while ignoring the small $\bar ff$ contributions to its width, we find
$\Gamma_{\st} \simeq 5\, (65)\,\gev$ for $M_{\st} = 300\,(600)\,\gev$. These
widths assume $M_{\tpi} = 0.55 \,M_{\st}$. They are roughly half this large
for $M_{\tpi} = 0.65\,M_{\st}$. The width of a 300 (600)~GeV standard-model
Higgs is about twice this large, 9 (125)~GeV.

\begin{figure}[!t]
 \begin{center}
\includegraphics[width=3.00in, height=3.00in]{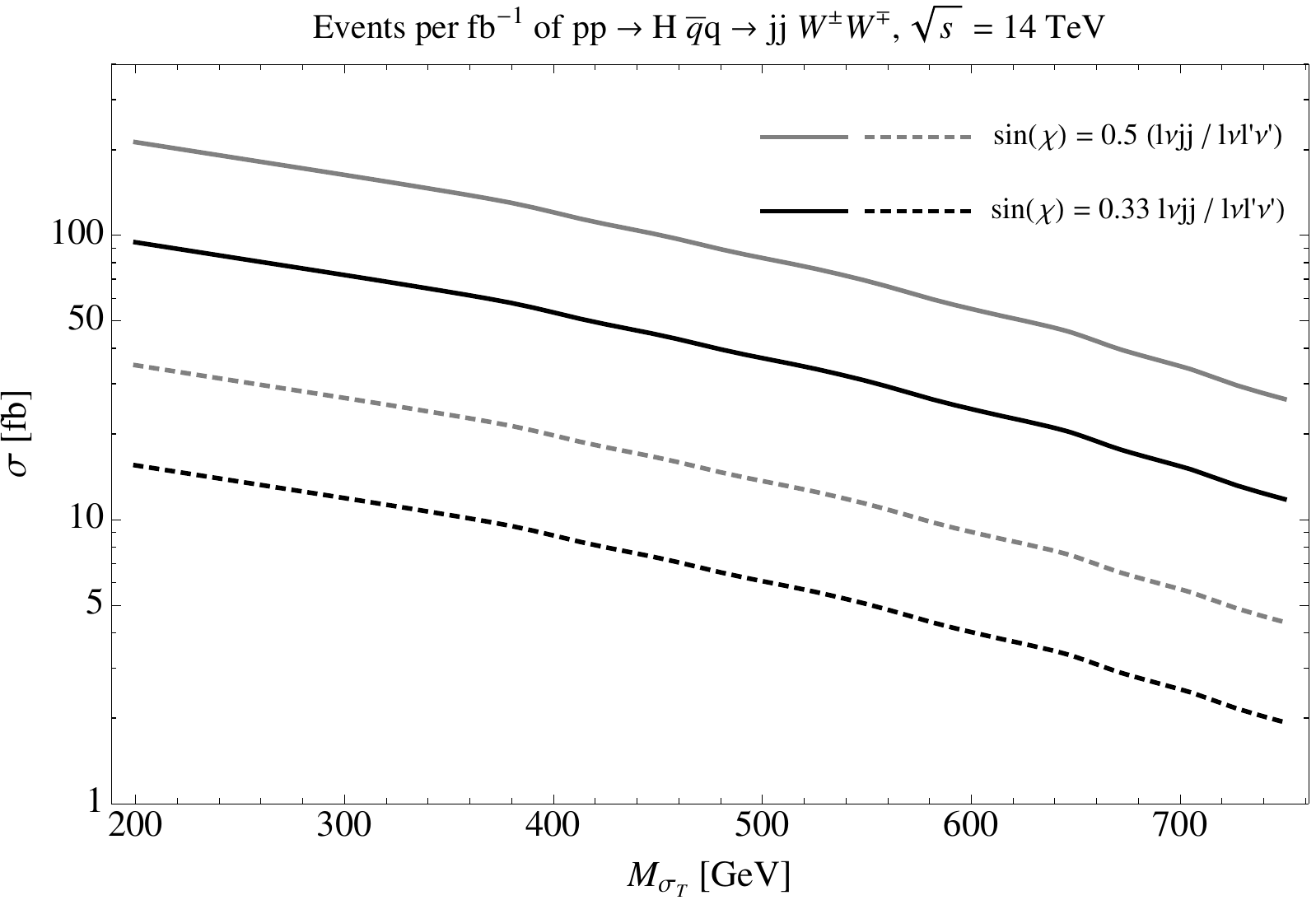}
\includegraphics[width=3.00in, height=3.00in]{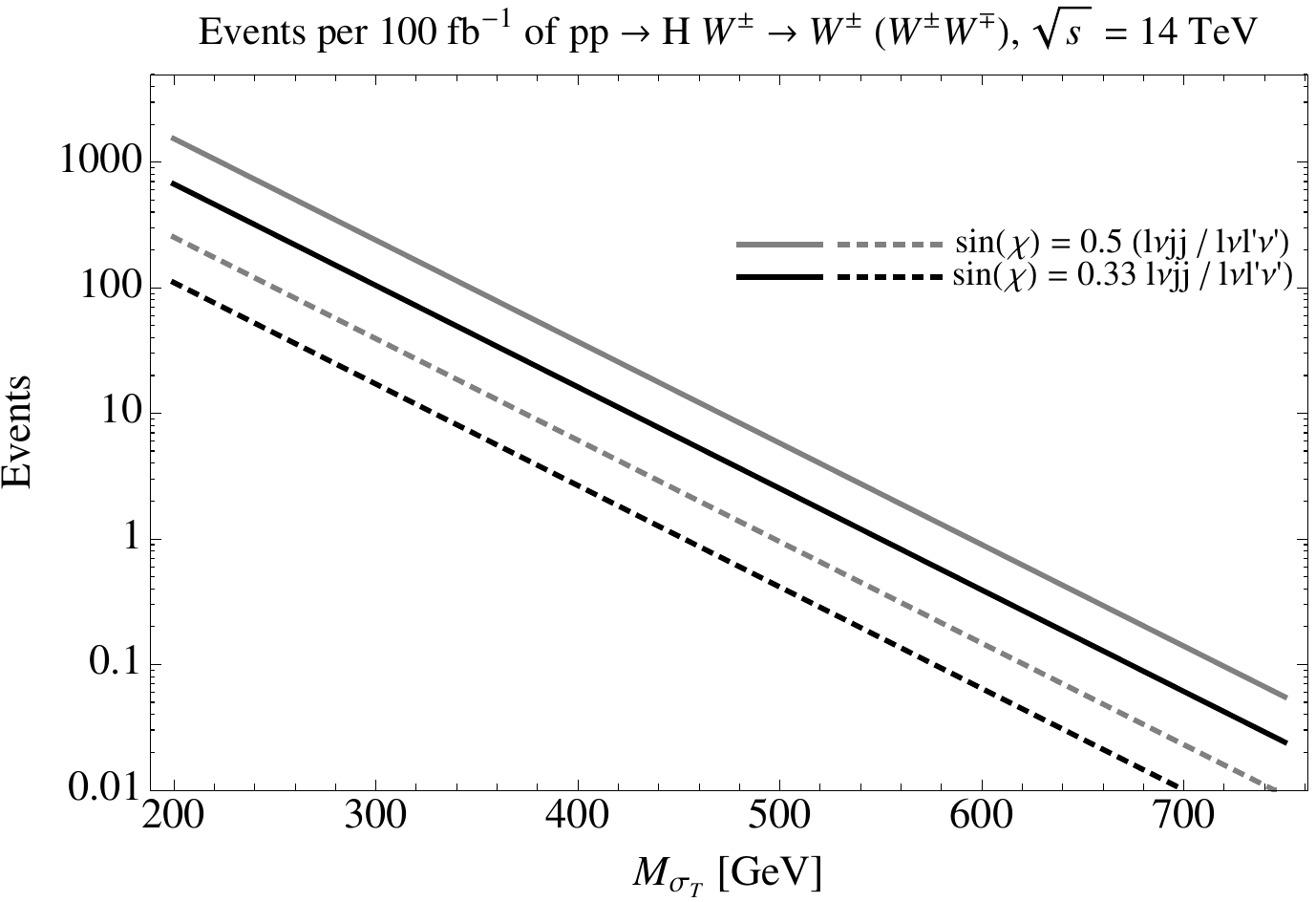}
\caption{The $\st \ra W^+W^-$ production rates in $pp$ collisions at
  $\sqrt{s} = 14\,\tev$ for $B(\st \ra W^+W^-) = 2/3$ and $\CK =1$ in
  Eq.~(\ref{eq:CK}); $\sin\chi = 1/3$ (black) or 1/2 (gray). Left: Weak boson
  fusion rates at $\int \CL dt = 1\,\ifb$. Right: Associated production
  $W^\pm \ra W^\pm + \st$ at $100\,\ifb$. Leptons include $e$ and
  $\mu$.\label{fig:sigma_rates}}
 \end{center}
 \end{figure}

 The VBF and associated production rates of $\st$ in $pp$ collisions at
 $14\,\tev$ and decay to $W^+W^-$ in the semileptonic $jj\ell\nu$ and
 leptonic $\ell\nu\ell'\nu'$ modes are shown in Fig.~\ref{fig:sigma_rates}.
 Here, $\ell,\ell' = e,\mu$. The upper limit $B(\st \ra W^+W^-) = 2/3$ was
 assumed. We show $\st$ decay branching ratios in Fig.~\ref{fig:sigma_BR} for
 $M_{\tpi}/M_{\st} = 0.55$ and 0.65.
 
 The raw VBF rates to $WW$ are large, but tagging the energetic forward jets
 is essential. The cleanest $WW$ mode is the leptonic one. The possibility of
 discovering
 the standard-model Higgs boson in this way, including a Higgs in the mass
 range of interest to us, was studied in Ref.~\cite{Cranmer:2004ys}.
 Rescaling the signal significances found there to a luminosity of
 $300\,\ifb$ yields Fig.~\ref{fig:WWlepsig}. This result is not promising for
 $\st$ discovery.
 
 A less explored and more promising possibility is VBF production followed by
 the semileptonic
 modes of $WW$ and $ZZ$. A complication for the semileptonic mode is that
 central jets must be vetoed to suppress $t\bar t + \jets $ and $W/Z + \jets$
 backgrounds while retaining enough hadronic activity to reconstruct the
 hadronically decaying gauge boson. This issue was addressed in CMS~Note
 2001/050, which studied VBF production of a heavy Higgs ($M_H = 300$ and
 $600\,\gev$) at the LHC with $\int \CL dt = 30\,\ifb$ at $\CL =
 10^{33}\,\cm^{-2}\,{\rm s}^{-1}$. The note found a central-jet veto
 effective in suppressing $\bar tt$ and a $W$-mass cut on $W\ra jj$
 suppressed $W/Z + \jets$. Translating their results to $\st$-production
 rates at a luminosity of $300\,\ifb$, we expect $S/\sqrt{S+B} = 0.7$ (1.6)
 for $M_{\st} = 300\,\gev$, $\sin\chi = 1/3$ and $M_{\tpi}/M_{\tpi} = 0.55$
 (0.65). The corresponding significances are 2.9 (4.7) for $\sin\chi = 1/2$.
 They are 1.5~to 2~times smaller than these for $M_{\st} = 600\,\gev$.

\begin{figure}[!ht]
 \begin{center}
\includegraphics[width=3.00in, height=3.00in]{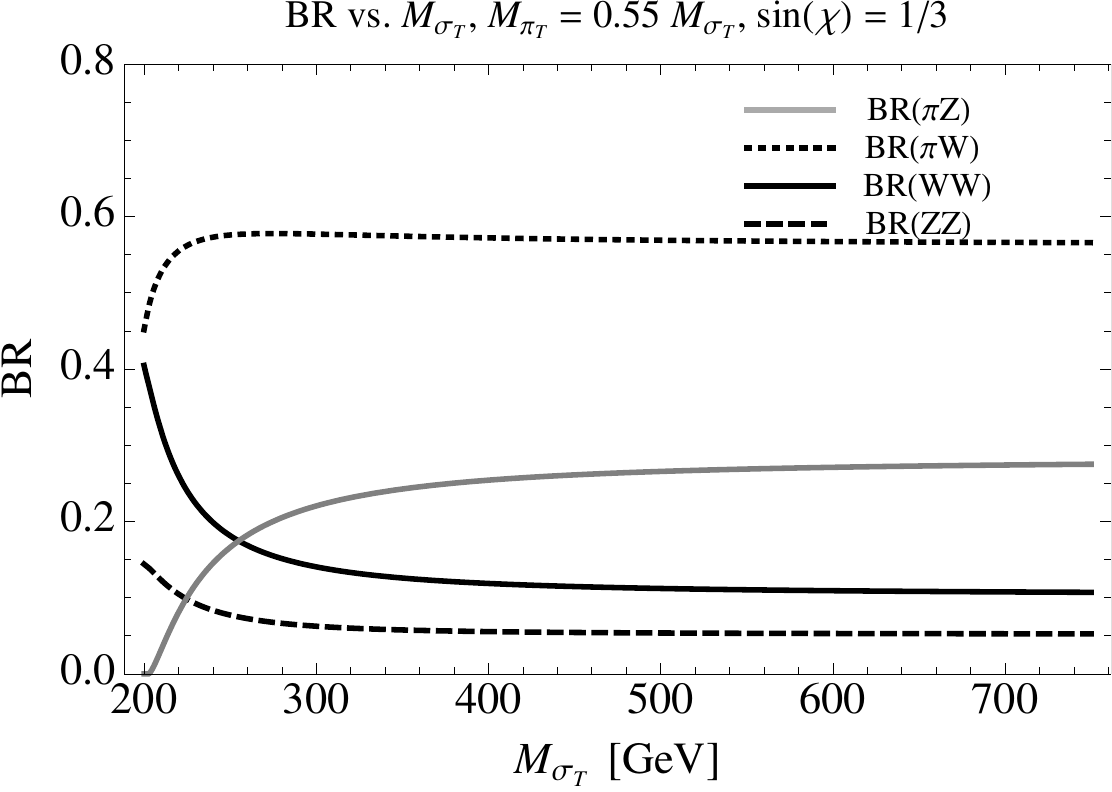}
\includegraphics[width=3.00in, height=3.00in]{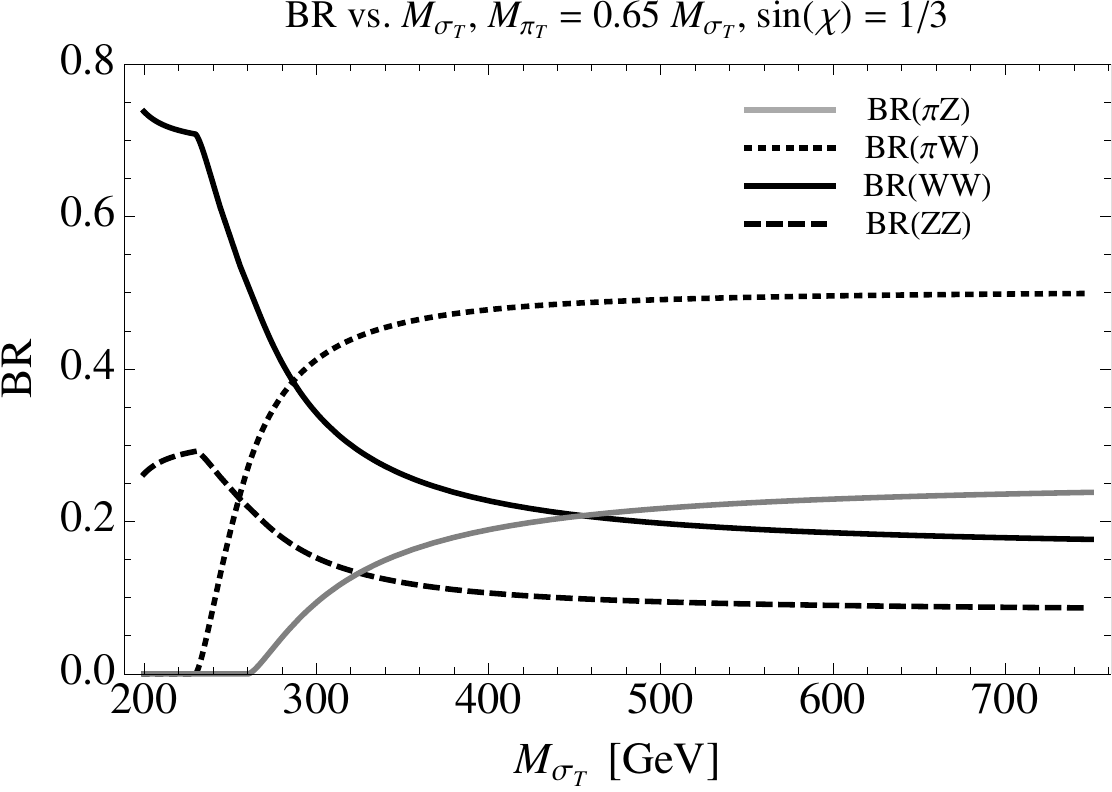}
\caption{The branching ratios for $\st \ra W^+ W^-$ (black), $Z^0 Z^0$
  (black dash), $\tpipm W^\mp$ (gray dash), and $\tpiz Z^0$ (gray) for
  $M_{\tpi}/M_{\st} = 0.55$ (left) and 0.65 (right). From in
  Eq.~(\ref{eq:szprod}) with $\CK=1$.\label{fig:sigma_BR}}
 \end{center}
 \end{figure}

 This CMS study was traditional in that it did not employ recently developed
 jet substructure techniques; see e.g., Refs.~\cite{Brooijmans:2008se,
   Butterworth:2008iy,Kaplan:2008ie,Thaler:2008ju,Plehn:2009rk,Katz:2010mr}. If a
 hadronically decaying gauge boson (more generally, any resonance) is
 sufficiently boosted, its daughter partons and their corresponding final
 state radiation can be captured within a single ``fat'' jet with radius $R
 \sim 1.2 - 1.5$. This fat jet will have high mass, $\sim M_{W,Z}$ and
 contain interior structure --- two subjet hotspots. Jet substructure
 techniques are easy to incorporate into VBF analyses. After tagging the
 forward jets, the remaining hadronic activity can be grouped into fat jets
 and analyzed for substructure. Then, once a hadronic gauge boson is
 identified, a jet veto can be applied to the remaining hadronic energy to
 further suppress $\bar t t + \jets$, etc.
 
One problem facing substructure techniques is that the large jet area
captures a lot of contamination from initial state radiation and the
underlying event --- energy not associated with the resonance. However, in
VBF there is no color information exchanged between the initial
quarks~\cite{Bjorken:1991xr, Bjorken:1992er, Barger:1994zq}.  This
makes VBF events less prone to these effects and therefore well-suited to
the use of substructure.
 
In a pioneering paper on jet substructure~\cite{Butterworth:2002tt}, the
method was applied to the search for strongly-interacting $WW$ resonances
produced via VBF. The analysis focused on discovering scalar and vector
resonances with masses greater than $1\ \tev$. We cannot simply recycle the
backgrounds of that analysis because it combined substructure and VBF cuts
with hard kinematic cuts on the $p_T$ of the reconstructed $W$ that
resonances like $\st$ in the $\sim 300$--$700\ \gev$ range will not pass.
Reference~\cite{Butterworth:2002tt} concluded that very heavy resonances
could be discovered and their decay angular distributions studied with
$100\,\ifb$ of LHC data. We expect a similar conclusion for a lighter, more
weakly interacting resonance. The analysis in Ref.~\cite{Butterworth:2002tt}
was done at the particle level, meaning that showering and hadronization were
included, but no detector effects other than fiducial volume cuts and
rudimentary particle ID efficiencies were applied. A detailed study
incorporating more realistic detector effects is needed.

\begin{figure}[!ht]
 \begin{center}
\includegraphics[width=3.00in, height=3.00in]{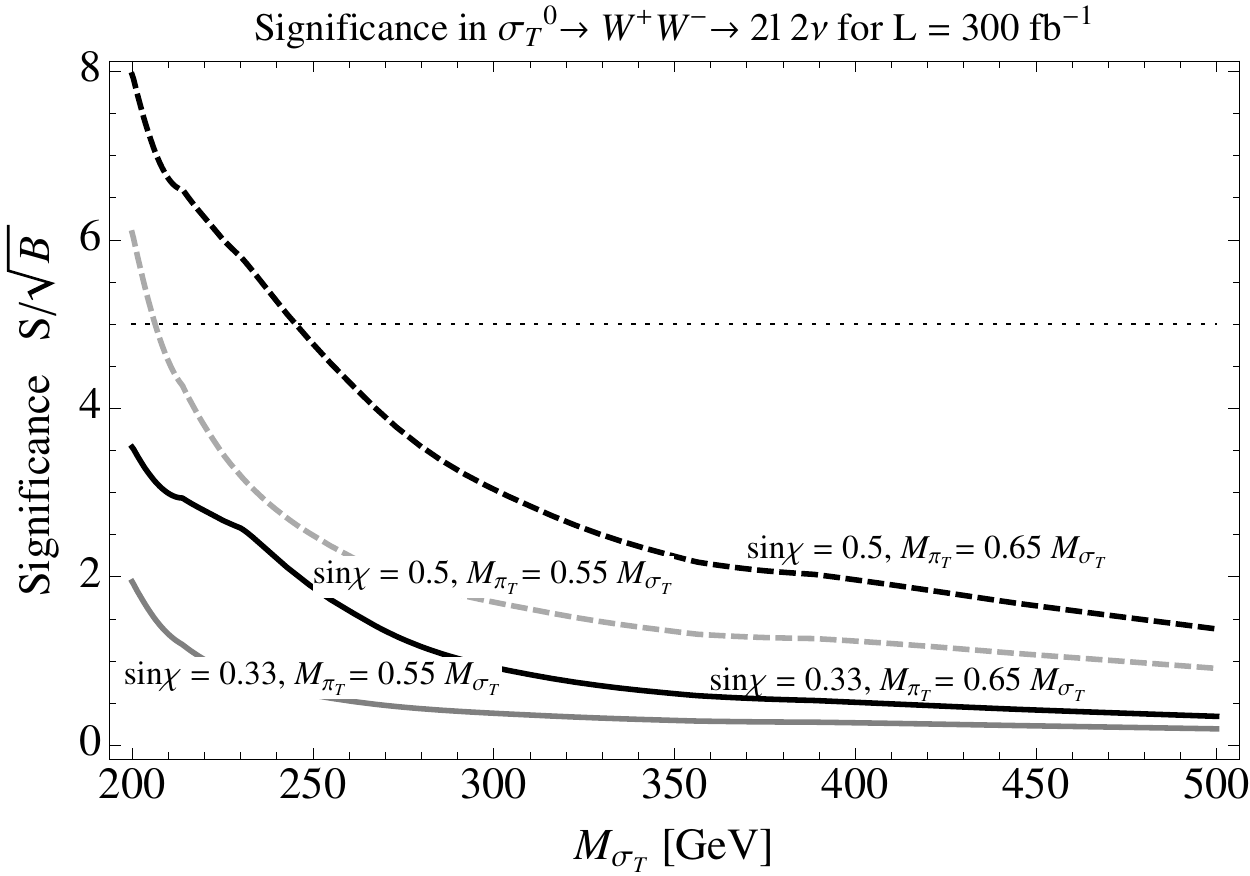}
\caption{Significance of $\st \ra W^+W^- \ra \ell\nu\ell'\nu'$ in pp
  collisions a $\sqrt{s} = 14\,\tev$ and $\int \CL dt = 300\,\ifb$ and for
  the indicated values of $\sin\chi$ and $M_{\tpi}/M_{\st}$; adapted from
  Ref.~\cite{Cranmer:2004ys}\label{fig:WWlepsig}}
 \end{center}
 \end{figure}

 The $\st \ra ZZ$ process suffers from the small $Z \ra \ellp \ellm$
 branching ratio and a production cross section half as large as for $WW$,
 but has the advantages of no missing $E_T$ and and no $W\ra \ell
 \nu$ reconstruction ambiguity. In Ref.~\cite{Hackstein:2010wk}, jet substructure
 techniques were applied to the decays of heavy Higgs bosons into $ZZ
 \rightarrow jj\ell^+\ell^-$.  Both gluon fusion and VBF production modes
 were studied and saw promising results. However, this analysis focused on
 substructure and did not combine hadronic $Z$ identification with the usual
 VBF selections and cuts, particularly on forward jets. Again, a more
 detailed and VBF-specific study is needed.
 
We have argued that the spectrum of low-scale technicolor has a relatively
light and relatively narrow $0^+0^{++}$ scalar, $\st$. Like a heavy Higgs
boson, $\st$ decays mainly to $WW$ and $ZZ$ and it is produced at the LHC via
$WW$ and $ZZ$ fusion, albeit at a rate suppressed by $\sin^2\chi \simeq
F_1^2/F_\pi^2 \sim 0.1$. The most promising final states are the semileptonic
ones in which one $W/Z$ decays hadronically. Tagging the forward jets of the
fusion process and using jet substructure to identify the hadronic $W$ or $Z$
decay may make $\st$ discovery possible, although it seems likely that a
luminosity of several $100\,\ifb$ will be required.  Detailed,
detector-specific simulations are necessary to decide this question, one we
think is well worth answering. If a standard-model-like Higgs boson is not
found at the LHC with luminosities typical of gluon fusion, it will be
important to determine whether any scalar exists. If the light technivectors
$\tro$, $\tom$, $\ta$ are found and they have nearly equal masses, the
discovery of $\st$ is then important to understanding the spectroscopy of
low-scale technicolor.

{\em Acknowledgments:} This work has been supported in part by the
U.S.~National Science Foundation under Grant~PHY-0905283-ARRA~(AD), the
U.S.~Department of Energy under Grant~DE-FG02-91ER40676~(KL), and Fermilab
operated by Fermi Research Alliance, LLC, U.S.~Department of Energy
Contract~DE-AC02-07CH11359 (AM). AD's research was also supported in part by
the CERN Theory Group and he thanks CERN for its hospitality. KL's research
was supported in part by Laboratoire d'Annecy-le-Vieux de Physique Theorique
(LAPTH) and the CERN Theory Group and he thanks LAPTH and CERN for their
hospitality. We gratefully acknowledge helpful discussions with Eric Pilon,
Maria Spiropulu, Brock Tweedie and Chris Vermilion.

\vfil\eject

\bibliography{LS-LSTC}
\bibliographystyle{utcaps}
\end{document}